\documentclass[letter]{article}
\usepackage[margin=1in]{geometry}
\usepackage{amsfonts,amsmath, amsthm, amssymb}
\usepackage{mathtools}
\usepackage{enumitem}
\usepackage{verbatim}
\usepackage{cancel}
\usepackage{color}
\usepackage{mathrsfs}
\usepackage{pdfsync}
\usepackage{extarrows}
\usepackage{comment}
\usepackage{lipsum}
\usepackage{physics}
\usepackage{sectsty}
\usepackage{setspace}
\usepackage{relsize}
\usepackage{graphicx}
\usepackage{caption}
\usepackage{subcaption}
\usepackage{bm}
\usepackage[english]{babel}

\begin{document}
\title{\vspace{-2.2em} Effects of head modeling errors on the spatial frequency representation of MEG}
\author{Wan-Jin Yeo$^{1,2}$, Eric Larson$^{2}$, Joonas Iivanainen$^{3}$, Amir Borna$^{3}$, Jim McKay$^{4}$, \\ Julia Stephen$^{5}$, Peter Schwindt$^{3}$, Samu Taulu$^{1,2}$ \\
\footnotesize{$^1$ Department of Physics, University of Washington, Seattle, WA 98195, United States}\\
\footnotesize{$^2$ Institute for Learning and Brain Sciences, University of Washington, Seattle, WA 98195, United States} \\
\footnotesize{$^3$ Sandia National Laboratories, Albuquerque, NM 87123, United States} \\
\footnotesize{$^4$ Candoo Systems Inc., Port Coquitlam, BC V3C 5M2, Canada} \\
\footnotesize{$^5$ The Mind Research Network, Albuquerque, NM 87106, United States}
}
\date{\today}
\maketitle

\begin{abstract}
Optically-pumped magnetometers (OPM) -- next-generation magnetoencephalography (MEG) sensors -- may be placed directly on the head, unlike the more commonly used superconducting quantum interference device (SQUID) sensors, which must be placed a few centimeters away. This allows for signals of higher spatial resolution to be captured, resulting in potentially more accurate source localization. In this paper, we show that in the noiseless and high signal-to-noise ratio (SNR) case of approximately $\geq 6$~dB, inaccuracies in boundary element method (BEM) head conductor models (or equivalently, inaccurate volume current models) lead to increased signal and equivalent current dipole (ECD) source localization inaccuracies when sensor arrays are placed closer to the head. This is true especially in the case of deep and superficial sources where volume current contributions are high. In the noisy case however, the higher SNR for closer sensor arrays allows for an improved ECD fit and outweighs the effects of head geometry inaccuracies. This calls for an increase in emphasis in head modeling to reduce inverse modeling errors, especially as the field of MEG strives for closer sensor arrays and cleaner signals. An analytical form to obtain the magnetic field errors for small perturbations in the BEM head geometry is also provided.
\end{abstract}

\section{Introduction}

Magnetoencephalography (MEG) is a non-invasive neuroimaging modality that provides spatiotemporal estimates of brain activity \cite{hamalainen, hari, ilmoniemi}. These estimates are based on inverse modeling, i.e., inferring the distribution of electric current in brain tissue based on a measurement of the associated magnetic field. Until recently, the only practical sensor type sensitive enough for MEG measurements has been the superconducting quantum interference device (SQUID), which requires cryogenics in order to maintain the superconducting state of the sensors. The necessary thermal insulation between the sensors, which must be held in a dewar containing liquid helium, and the surface of the head creates a minimum gap of at least 2~cm between the scalp and the sensors. However, reducing the measurement distance is imperative to further improve spatial resolution of MEG as the magnetic fields decay rapidly as a function of distance. Fortunately, recent developments in sensor technology, especially in the domain of optically pumped magnetometers (OPM), have made on-scalp MEG possible \cite{borna, borna1, boto, hill, iivanainen1, knappe}. The reduced proximity between the sensors and the brain sources has the potential to significantly improve the spatial resolution of MEG, since the detected signal strength and spatial complexity are expected to increase.

Brain sources are typically modeled as having a primary current and volume current. The primary current is where neural activity occurs and is the portion of the current that is of primary interest. In many cases, the primary current can be modeled as a current dipole, especially for focal sources. The volume current is the passive return current that closes the current loop inside the head. Because volume currents are mathematically equivalent to surface currents on the conductivity boundaries of closed, piece-wise homogeneous conductor models of the head \cite{geselowitz}, any inaccuracies in head geometries that are involved in forward signal calculations will inaccurately account for the volume current, resulting in an incorrect forward model. One such forward calculation method is the boundary element method (BEM), which typically uses triangulated, decimated surface meshes for the head model so that the electric potential, which is implicitly defined and difficult to determine in the continuous case, may be more easily calculated through some assumptions, such as uniform conductivity within compartments \cite{makinen,phillips,stenroosog}. 

For OPMs that can potentially detect signals with higher spatial resolutions due to their decreased sensor-to-array distance, higher spatial frequency contributions from primary and volume currents are expected to be detected. Here, we investigate the importance of BEM triangle mesh accuracy in forward modelling of MEG signals as a function of sensor array distances by perturbing vertices of the mesh to imitate mesh inaccuracies. We also investigate which spatial frequencies suffer from the greatest errors as a function of sensor array distance. We used the simplest and most straightforward BEM methods, the constant collocation (CC) and linear collocation (LC) approaches, to illustrate this effect. The simplicity of CC BEM has the added advantage of being able to easily illustrate the general method one may use to find analytical forms of the errors, as shown in Appendix \ref{appendix_analytical}. Then, we illustrate the effects of the inaccurate forward signals in equivalent current dipole (ECD) source localization fits. We show that in the noiseless case, increased signal error due to BEM errors result in less accurate source localization, especially for deep sources. However, in the presence of noise, the increased SNR due to closer sensor array distances to the head generally still improves the source localization.

There are many sources of errors for BEM, which may be broadly categorized as anatomical modeling errors and/or numerical errors. For example, the effect of conductivity errors on forward signals have been studied in \cite{haueisen, stenroosmne, stenrooscsf,vallaghe}, and different mesh triangulation methods and basis choices to estimate the potentials within each triangle may result in numerical errors that affect accuracy of forward calculations \cite{meijs,schlitt}. Studies on head geometry errors, especially focusing on the skull, have also been conducted in for example \cite{dannhauer, lanfer}. An overview of many of these errors can be found in \cite{ferguson1}. In this paper, we specifically investigate the effect of random vertex perturbations from an ideal spherical head model on the resulting signal. One other paper has explored head model errors from ideal spherical models as well, but with the assumption that the errors are sufficiently smooth so that they can be described with a spherical harmonic expansion accurately \cite{nolte}. Since our approach accounts for individual vertex perturbations, it may offer a solution to more precise adjustments/corrections to discretized head models.

We first give an overview of BEM used in MEG signal forward calculations, with emphasis on the CC approach. Then, we discuss our results for the effects of BEM head geometry inaccuracies on the signal vectors for varying sensor array and source distances. We also analyzed the errors in the spatial frequency domain via a decomposition into the Signal Space Separation (SSS) basis with varying $l$-degree truncation choices. Finally, we performed ECD fits with the inaccurate forward calculated signals to determine their effects on source localization in both noiseless and noisy sensor cases. In the appendix, analytical forms that one may use as an alternative way to calculate signal perturbations are also provided.

\section{Boundary Element Method overview} \label{sec:BEM2}

\subsection{Geselowitz' formula}

It is common practice in MEG and EEG modeling to express the total current as a superposition of two components, i.e.,
\begin{equation}\label{J_division}
\mathbf{J} \left( \mathbf{r}' \right) = \mathbf{J}^P \left( \mathbf{r}' \right) + \mathbf{J}^v \left( \mathbf{r}' \right),
\end{equation}
where 
$\mathbf{J}^P \left( \mathbf{r}' \right)$ represents the physiologically interesting primary current and $\mathbf{J}^v \left( \mathbf{r}' \right)$ is the associated passive volume current component that completes the loop of electric current in the brain tissue. On the basis of the Coulomb force, the passive volume current is of the form 
\begin{equation}\label{Jv_expression}
    \mathbf{J}^v \left( \mathbf{r}' \right) = \sigma\mathbf{E} = -\sigma\nabla V,
\end{equation}
where $\sigma$, $\mathbf{E}$, and $V$ are the electric conductivity, the electric field, and the electric potential, respectively. The second equality in equation \eqref{Jv_expression} is valid when the time derivative of the magnetic field $\mathbf{B}$ is insignificant, which is the case in MEG and EEG \cite{hamalainen}. 
If we assume a piecewise homogeneous conductor head model with $N_S$ conductivity boundary surfaces, the magnetic field due to $\mathbf{J} \left( \mathbf{r}' \right)$ is given by \textit{Geselowitz' formula} \cite{geselowitz,hamalainen}
\begin{equation}\label{mainB}
    \mathbf{B} \left(\mathbf{r} \right) = \mathbf{B}_0 \left(\mathbf{r} \right) + \frac{\mu_0}{4 \pi} \sum_{l=1}^{N_S} \left(\sigma_l^- - \sigma_l^+ \right) \int_{S_l'} V \left(\mathbf{r}'\right) \frac{\mathbf{r}-\mathbf{r}'}{\abs{\mathbf{r}-\mathbf{r}'}^3} \times d \mathbf{S}'_{l},
\end{equation}
where 
\begin{equation}\label{B0}
    \mathbf{B}_0 \left(\mathbf{r} \right) = \frac{\mu_0}{4 \pi} \int_{v'} \mathbf{J}^P \left( \mathbf{r}' \right) \times \frac{\mathbf{r}-\mathbf{r}'}{\abs{\mathbf{r}-\mathbf{r}'}^3} \ dv'
\end{equation}
is the contribution by primary currents, $v'$ is the total head volume, and $\sigma_l^-$ and $\sigma_l^+$ denote conductivities of the inner and outer regions relative to $S_l$. The second term on the right hand side of \eqref{mainB} is the contribution by volume currents, which we will denote as $\mathbf{B}_{vol}$. We need the electric potential $V$ in order to calculate $\mathbf{B}_{vol}$ and hence the total magnetic field. Below we review the expressions to obtain discrete values of $V$ on the boundaries that we may thus use to approximate and discretize $\mathbf{B}$.

\subsection{Discretization of the electric potential field}

Here, we give an overview of the CC BEM approach. The LC requires only a slight modification, as will be pointed out. More comprehensive reviews may be found in \cite{demunck, ferguson2, phillips, stenroosall, stenroosog}.

The electric potential $V(\mathbf{r})$ for a field point $\mathbf{r}$ on the $k$\textsuperscript{th} surface $S_{k}$ is given intrinsically by \cite{hamalainen, phillips}
\begin{equation}\label{mainV}
    V \left( \mathbf{r} \right) = V_\infty \left( \mathbf{r} \right) - \frac{1}{2\pi} \sum_{l=1}^{N_S} \frac{\sigma_l^- - \sigma_l^+}{\sigma_k^- + \sigma_k^+} \int_{S_{l}'} V\left( \mathbf{r}' \right) \frac{\mathbf{r}-\mathbf{r}'}{\abs{\mathbf{r}-\mathbf{r}'}^3} \cdot d \mathbf{S}'_{l}
\end{equation}
where
\begin{equation}\label{Vinf}
    V_\infty \left( \mathbf{r} \right) =  \frac{1}{2 \pi \left(\sigma_k^- + \sigma_k^+ \right)} \int_{v'} \mathbf{J}^P \left( \mathbf{r}' \right) \cdot \frac{\mathbf{r}-\mathbf{r}'}{\abs{\mathbf{r}-\mathbf{r}'}^3} \ dv'.
\end{equation}
Like in the $\mathbf{B}$ field case, eq. \eqref{Vinf} describes the primary current contribution to the potential whereas the surface integral is the volume current contribution. The primary current contribution is easily obtained if we have prescribed $\mathbf{J}^P$. In particular, if we let $\mathbf{J}^P$ be a collection of current dipoles with positions represented by delta functions, the integral collapses to a simple form. For $N$ dipoles $\mathbf{J}_n^P = \mathbf{Q}_n \delta(\mathbf{r}-\mathbf{r}_n)$ where $n = 1,\dots,N$, we have 
\begin{equation}
    V_\infty\left(\mathbf{r}\right) = \frac{1}{2\pi\left(\sigma_k^- + \sigma_k^+ \right)} \sum_{n=1}^{N} \mathbf{Q}_n \cdot \frac{\mathbf{r}-\mathbf{r}_n}{\abs{\mathbf{r}-\mathbf{r}_n}^3}.
\end{equation}
As for the volume current contribution, we may discretize each surface $S_l$ into $N_l$ triangles. Then, \eqref{mainV} can be written as
\begin{equation}
    V\left(\mathbf{r}\right) = V_\infty \left(\mathbf{r}\right) - \frac{1}{2\pi} \sum_{l=1}^{N_S} \frac{\sigma_l^- - \sigma_l^+}{\sigma_k^- + \sigma_k^+} \sum_{m=1}^{N_l} \int_{\Delta_l^m} V\left( \mathbf{r}' \right) \frac{\mathbf{r}-\mathbf{r}'}{\abs{\mathbf{r}-\mathbf{r}'}^3} \cdot d \mathbf{S}_{\Delta_l^m}'
\end{equation}
where $\Delta_l^m$ is the $m$\textsuperscript{th} triangle of surface $S_l$.

If we have chosen a large enough $N_l$ such that the triangle areas are small, we may reasonably make some assumptions about the behavior of the potential $V$ within each triangle. In turn, $V$ can be estimated with some basis and weight functions defined relative to parameters of relevant triangles, independent of $\mathbf{r}$. The potential $V$ can then be explicitly defined and solved for. Many such approximations exist, including using constant or linear basis with collocation or Galerkin weighting \cite{stenroosall}. Higher-degree basis functions have been considered as well \cite{gencer}. 

In this paper, we use BEM as a tool to calculate the errors of volume current contributions due to closer sensors. Since we are not actually concerned with the accuracy between different approximation methods but rather the overall behaviour of the resulting signal due to head geometry errors, the most straightforward approximations which are the CC and LC approaches suffice for our purposes. We present the CC case in the next section which assumes a constant potential within each triangle. The LC approach approximates the potential within each triangle as a linear function via an interpolation from the potentials at the three vertices; see \cite{demunck,stenroosall,stenroosog}. The simplicity of the CC approach allows us to outline an analytical method one may use to find errors for small boundary perturbations as well (Appendix \ref{appendix_analytical}).

\subsection{Linearization of electric potentials}

First, we may assume that triangles are small enough such that the potential is constant in each triangle, i.e., $V(\mathbf{r}') \approx V(\mathbf{c}_l^m)$ when $\mathbf{r}' \in \Delta_l^m$, where $\mathbf{c}_l^m$ is the centroid of $\Delta_l^m$. This allows us to pull the potential term out of the integral, and we get 
\begin{equation}
    \int_{\Delta_l^m} V\left( \mathbf{r}' \right) \frac{\mathbf{r}-\mathbf{r}'}{\abs{\mathbf{r}-\mathbf{r}'}^3} \cdot d \mathbf{S}_{\Delta_l^m} \approx V\left( \mathbf{c}_l^m \right) \int_{\Delta_l^m} \frac{\mathbf{r}-\mathbf{r}'}{\abs{\mathbf{r}-\mathbf{r}'}^3} \cdot d \mathbf{S}_{\Delta_l^m}'.
\end{equation}
Notice that the integral on the right hand side is now simply the solid angle spanned by the triangle $\Delta_l^m$ from the observation point $\mathbf{r}$; let us denote it as $\Omega_l^m \left(\mathbf{r}\right)$. If we let $\mathbf{r}_{i0} \equiv \mathbf{r}_i - \mathbf{r}$, $i = 1,2,3$ be the three vertices of the triangle relative to $\mathbf{r}$, and let $r_{i0}$ be their lengths, we may equivalently express each $\Omega_l^m$ as \cite{oosterom}
\begin{equation}\label{solidangle}
    \Omega_l^m = 2\arctan \left[ \frac{\mathbf{r}_{10} \cdot \left(\mathbf{r}_{20} \times \mathbf{r}_{30} \right)}{r_{10} r_{20} r_{30} + \left(\mathbf{r}_{10} \cdot \mathbf{r}_{20} \right)r_{30} + \left(\mathbf{r}_{30} \cdot \mathbf{r}_{10} \right)r_{20} + \left(\mathbf{r}_{20} \cdot \mathbf{r}_{30} \right)r_{10}}\right].
\end{equation}

Let $\mathbf{r}$ coincide with centroids of the triangles as well. Then, all the potential terms of \eqref{mainV} are discretized at the same locations and it can now be compactly written as 
\begin{equation}
    V \left(\mathbf{c}_k^i \right) = V_\infty \left(\mathbf{c}_k^i\right) - \frac{1}{2\pi} \sum_{l=1}^{N_S} \sum_{m=1}^{N_l} \frac{\sigma_l^- - \sigma_l^+}{\sigma_k^- + \sigma_k^+} V\left(\mathbf{c}_l^m \right)  \Omega_l^m \left(\mathbf{c}_k^i\right).
\end{equation}
In matrix form, this looks like 
\begin{equation}\label{matrixeqn}
    \begin{bmatrix}
    \mathbf{V}_1 \\
    \vdots \\
    \mathbf{V}_{N_S}
    \end{bmatrix} 
    = \begin{bmatrix}
    \mathbf{V}_{\infty,1}  \\
    \vdots \\
    \mathbf{V}_{\infty,N_S}
    \end{bmatrix} 
    + \begin{bmatrix}
    \mathbf{G}_{1,1} && \cdots && \mathbf{G}_{1,N_S} \\
    \vdots && \ddots && \vdots \\
    \mathbf{G}_{N_S,1} && \cdots && \mathbf{G}_{N_S,N_S} \\
    \end{bmatrix}
    \begin{bmatrix}
    \mathbf{V}_1 \\
    \vdots \\
    \mathbf{V}_{N_S}
    \end{bmatrix} 
\end{equation}
where
\begin{equation} \label{G}
    G_{k,l}^{i,m} =  - \frac{1}{2\pi} \frac{\sigma_l^- - \sigma_l^+}{\sigma_k^- + \sigma_k^+} \Omega_l^m \left(\mathbf{c}_k^i \right).
\end{equation}
Note that the $\mathbf{G}$ matrix is dependent only on the geometry and conductivities of the conductor, and it is also the only term that depends on the boundary geometry. This means that for each different source configuration in the same head model, $\mathbf{G}$ needs to be calculated only once, whereas the primary current contribution $\mathbf{V}$ needs to be recalculated.

\subsection{Matrix deflation}

If we write the matrix equation \eqref{matrixeqn} as $\mathbf{V} = \mathbf{V}_\infty + \mathbf{G} \mathbf{V}$, then we have
\begin{equation}\label{undeflated}
    \left(\mathbb{I}-\mathbf{G} \right) \mathbf{V} = \mathbf{V}_\infty
\end{equation}
where $\mathbb{I}$ is the identity matrix. It is seemingly straightforward to solve for $\mathbf{V}$ by taking the inverse of $(\mathbb{I}-\mathbf{G})$, but $(\mathbb{I}-\mathbf{G})$ is actually non-invertible since it is rank-deficient; the electric potential has infinite number of solutions since it is defined up to an additive constant. This manifests from the fact that the fundamental equation we are trying to solve is the Poisson equation within the head
\begin{equation}
    \grad \cdot \left(\sigma \grad V \right) = \grad \cdot \mathbf{J}^p
\end{equation}
with Neumann boundary condition at each boundary (current continuity)
\begin{eqnarray}
    \sigma^+ \grad V \cdot \mathbf{n} = \sigma^- \grad V \cdot \mathbf{n}
\end{eqnarray}
where $\mathbf{n}$ is the outward-pointing unit surface normal. These equations are specified only up to the first derivative/gradient of $V$, hence $V$ is defined up to a constant.

We thus know that both $\mathbf{V}$ and $\mathbf{V} + k \mathbf{e}$, where $\mathbf{e} = (1,1,\dots,1)^T$ and $k$ is a nonzero constant, are solutions to the matrix equation. So, in addition to \eqref{undeflated}, we also have 
\begin{equation}
    \left(\mathbb{I}-\mathbf{G} \right) \left( \mathbf{V} + k \mathbf{e} \right) = \mathbf{V}_\infty.
\end{equation}
Subtracting \eqref{undeflated} from this yields
\begin{equation}
    \left(\mathbb{I}-\mathbf{G} \right) \mathbf{e} = 0
\end{equation}
which indicates that $\left(\mathbb{I}-\mathbf{G} \right)$ has a zero eigenvalue with associated eigenvector $\mathbf{e} \neq \mathbf{0}$, i.e. it is indeed singular. Equivalently,
\begin{equation}\label{deflatecond}
\mathbf{G} \mathbf{e} = \mathbf{e}.  
\end{equation}
This means that $\mathbf{G}$ has a unit eigenvalue with corresponding eigenvector $\mathbf{e}$. So, one way to avoid the singularity is to eliminate this unit eigenvalue; the standard way to do this is by \textit{deflation}, as follows.

First, assume the unit eigenvalue of $\mathbf{G}$ is simple \cite{lynn}. For any vector $\mathbf{a}$, we need to find a vector $\mathbf{c}$ with constant entries (not all necessarily the same) such that
\begin{eqnarray}
\mathbf{c}^T \mathbf{a} = 
\begin{cases}
    k \ \ \text{if} \ \mathbf{a} = k\mathbf{e} \\
    0 \ \ \text{otherwise.}
\end{cases}
\end{eqnarray}
The first case imposes the condition of defining a reference potential in some way. For example, if we pick $\mathbf{c}$ to have all the same entries, then it means we let the sum of all potentials over all boundaries to be zero \cite{hamalainenisa}. We may also pick just a few entries to be zero, corresponding to a possibly more meaningful reference potential; for example, Wilson terminals are used in electrocardiogram \cite{fischer}. The second case ensures that all eigenvalues of $\mathbf{G}' = (\mathbf{G} - \mathbf{e}\mathbf{c}^T)$ are equal to the eigenvalues of $\mathbf{G}$, except for the unit eigenvalue which is replaced by zero. This ensures the $\mathbf{G}'$ is non-singular and hence invertible by condition \eqref{deflatecond}. We may explicitly show this preservation of eigenvalues for $\mathbf{G}'$ as follows. Let $\lambda$ and $\mathbf{v}_e$ be eigenvalues and corresponding eigenvectors of $\mathbf{G}$. For $\mathbf{v}_e \neq \mathbf{e}$,
\begin{eqnarray}
    \mathbf{G}'\mathbf{v}_e = \left(\mathbf{G} - \mathbf{e} \mathbf{c}^T \right) \mathbf{v}_e = \mathbf{G}\mathbf{v}_e - \mathbf{e} \mathbf{c}^T \mathbf{v}_e = \lambda \mathbf{v}_e - \mathbf{e}0 = \lambda \mathbf{v}_e
\end{eqnarray}
hence eigenvalues are preserved. For $\mathbf{v}_e = k\mathbf{e}$,
\begin{eqnarray}
    \mathbf{G}'k\mathbf{e} = \left(\mathbf{G} - \mathbf{e} \mathbf{c}^T \right) k\mathbf{e} = \mathbf{G}k\mathbf{e} - \mathbf{e} \mathbf{c}^T k\mathbf{e} = k\mathbf{e} - \mathbf{e}k = 0.
\end{eqnarray}
So, we are now able to solve for $\mathbf{V}$ with the invertible deflated $(\mathbb{I} - \mathbf{G}')$, 
\begin{equation}
    \mathbf{V} = \left(\mathbb{I}-\mathbf{G} + \mathbf{e}\mathbf{c}^T \right)^{-1} \mathbf{V}_\infty.
\end{equation}

\subsection{Discretization of the magnetic field}

We now have a set of discrete potential solutions at the centroids of the triangles. If we discretize the $\mathbf{B}_{vol}$ integral in an identical manner as the electric potential case above, it allows us to get an approximation of the magnetic field at arbitrary field locations $\mathbf{r}$ directly from \eqref{mainB}. The magnetic field is given by
\begin{align}\label{Bfield_CC}
    \mathbf{B}\left(\mathbf{r} \right) \approx \mathbf{B}_0 \left( \mathbf{r} \right) + \frac{\mu_0}{4 \pi} \sum_{l=1}^{N_S} \left(\sigma_l^- -\sigma_l^+\right) \sum_{m=1}^{N_l} V\left(\mathbf{c}_l^m \right) \mathbf{\Omega}_l^m
\end{align}
where we have defined the ``vector solid angle''
\begin{align}
    \mathbf{\Omega}_l^m &\equiv \int_{\Delta_l^m} \frac{\mathbf{r}-\mathbf{r}'}{\abs{\mathbf{r}-\mathbf{r}'}^3}\times d\mathbf{S}_{\Delta_l^m}'.
\end{align}
The evaluation of $\mathbf{\Omega}_l^m$ has been done in \cite{demunck} via Stoke's Theorem, and is given by
\begin{equation} \label{vecomega}
    \mathbf{\Omega}_l^m = \sum_{i=1}^{3} \left(\gamma_{i-1} - \gamma_i \right) \mathbf{r}_i
\end{equation}
where
\begin{equation}
    \gamma_i \equiv -\frac{1}{\abs{\mathbf{r}_{i+1} - \mathbf{r}_i}} \cdot \ln \frac{r_i \abs{\mathbf{r}_{i+1} - \mathbf{r}_i} + \mathbf{r}_i \cdot \left(\mathbf{r}_{i+1} - \mathbf{r}_i \right)}{r_{i+1} \abs{\mathbf{r}_{i+1} - \mathbf{r}_i} + \mathbf{r}_{i+1} \cdot \left(\mathbf{r}_{i+1} - \mathbf{r}_i \right)}
\end{equation}
and $i = 1, 2, 3$ correspond to the three vertices of triangle $m$ on surface $S_l$. Also note that $\mathbf{r}_4 \equiv \mathbf{r}_1$ and $\mathbf{r}_0 \equiv \mathbf{r}_3$. 

This expression allows us to calculate the magnetic field easily, assuming useful indexing of vertices and triangles has been done in the process of surface discretization.

\subsection{Calculation of magnetic flux signals}

For forward calculation of the signal vectors, we calculate the magnetic flux through magnetometer pick-up loops. In reality, this pick-up loop setup corresponds only for SQUIDs; OPMs measure the volume integral of the magnetic field over a cylindrical sensing volume. However, we are interested primarily in how the reduced distance of OPM sensor arrays may affect the signal measured due to BEM head model errors, hence we do not consider volume integrals over the magnetic field here. 

For $N$ sensors, the magnetic field is discretized into $N$ channel readings of magnetic flux. The vectorization of these readings into an $N \times 1$ vector $\bm{\phi}$ is defined as the \textit{signal vector}. The \textit{signal space} (or \textit{signal}) is the $N$-dimensional vector space with elements being any possible signal vector. Within the context of BEM, the signal space may be defined as the space containing all possible signals one may obtain when doing a forward calculation using the triangulation of a perfectly accurate head model.

In the case of sensors measuring the magnetic flux through a surface specified by an area and a normal vector, the $j$\textsuperscript{th} element of $\bm{\phi}$ is given by
\begin{equation}
    \phi_{j} = \int_{S_{j}}\mathbf{B}(\mathbf{r})\cdot d\mathbf{S}_{j},
    \label{eq:phi_element}
\end{equation}
where $d\mathbf{S}_{j}$ represents an infinitesimal surface element on the sensor surface with unit normal $\mathbf{n}_{j}$. The calculation of \eqref{eq:phi_element} is commonly done by cubature approximation over the sensor area \cite{abramowitz}.

\section{Representation of perturbations in the signal and source space} \label{sec:BEM3}

\subsection{Additive perturbation of signal space}

From the BEM steps above, we see that any inaccurate head mesh models will lead to inaccurate forward calculations of the magnetic flux, since they correspond to perturbed triangle vertices and centroids. This in turn leads to inaccurate calculations of the potential, magnetic field and magnetic flux signal vector. These errors may be written as an additive perturbation since they are random and independent of the unperturbed quantities, 
\begin{align}
    \mathbf{V}' = \mathbf{V} + \delta \mathbf{V} \\ 
    \mathbf{B}' = \mathbf{B} + \delta \mathbf{B} \\
    \bm{\phi}' = \bm{\phi} + \delta\bm{\phi}
\end{align}
where the perturbations to each element of the flux signal vector are given by
\begin{equation}
    \delta\phi_{j} = \int_{S_{j}}\delta\mathbf{B}(\mathbf{r})\cdot d\mathbf{S}_{j},
    \label{eq:phi_element_perturb}
\end{equation}

Note that the BEM errors are only relevant to the forward modeling of the signal vectors; real recorded signal vectors by definition do not have BEM errors. In the context of (noiseless) BEM forward modeling, the goal is to set up a head model such that the calculated signal $\bm{\phi}'$ is as close to the (noiseless) recorded/true data $\bm{\phi}$ as possible. If information about the BEM geometry is perfect, then $\bm{\phi}'=\bm{\phi}$ with $\delta\bm{\phi} = 0$. We re-emphasize that the primary current contribution \eqref{B0} does not depend on the head model by Geselowitz's formula, thus all errors come from inaccurate volume contribution. In other words, $\delta \mathbf{B} = \delta \mathbf{B}_{vol}$ and hence $\delta \bm{\phi} = \delta \bm{\phi}_{vol}$.

As an aside in Appendix \ref{appendix_analytical}, we present an analytical approach to calculate the first-order perturbation contributions of $\delta \mathbf{V}$ and $\delta \mathbf{B}$. The subsequent calculation of the flux perturbation $\delta \bm{\phi}$ may be done numerically with cubature approximations \cite{abramowitz} or analytically \cite{yeo}.

\subsection{Quantification of signal reconstruction errors with subspace angle} \label{section:angles_ldeg}

A compact metric for quantifying the difference between the recorded/reference data and modeled/perturbed data is the angle between the corresponding signal vectors $\bm{\phi}$ and $\bm{\phi}'$ respectively. We may calculate their subspace angle $\theta$ by \cite{gunawan}
\begin{equation} \label{angle}
    \text{subspace}(\bm{\phi},\bm{\phi}') = \theta = \arccos \left( \frac{\lVert \text{Proj}_{\bm{\phi}} \bm{\phi}' \rVert}{\lVert \bm{\phi} \rVert} \right).
\end{equation}
If we decompose the signals into their different spatial frequency components via the SSS method \cite{taulu}, then the angle for individual different spatial frequency components can be calculated as well. In the SSS formalism, signal vectors are decomposed into their $l$ degree components via a vector spherical harmonic (VSH) expansion, and each spatial frequency component is conveniently characterized by an $l$ degree component. Higher $l$ degrees correspond to higher spatial frequencies, whereas lower $l$ degrees correspond to lower spatial frequency signal components. Let $\mathbf{S}$ be the basis matrix containing the VSH modes, $\mathbf{S}_{1:L}$ be the first $l=L$ degree portion of $\mathbf{S}$, and $\mathbf{x}_{1:L}$ be the corresponding coefficients/multipole moments of $\mathbf{S}_{1:L}$. The subspace angle $\theta_l$ specific to the cumulative spatial frequency bands from $l=1$ to $l=L$ is thus
\begin{equation} \label{anglel}
    \theta_{1:L} = \arccos \left( \frac{\lVert \text{Proj}_{\bm{\phi}_{1:L}} \bm{\phi}_{1:L}' \rVert}{\lVert \bm{\phi}_{1:L} \rVert} \right),
\end{equation}
where
\begin{equation}
    \bm{\phi}_{1:L} = \mathbf{S}_{1:L}\mathbf{x}_{1:L}
\end{equation}
and
\begin{equation}
    \bm{\phi}_{1:L}' = \mathbf{S}_{1:L}\mathbf{x}_{1:L}'.
\end{equation}
Given a measured signal vector $\bm{\phi}$, the multipole moments $\mathbf{x}_{1:L}$ must first be estimated by taking the pseudoinverse of $\mathbf{S}_{1:L}$,
\begin{equation}
    \mathbf{x}_{1:L} \approx \mathbf{S}_{1:L}^{\dagger} \bm{\phi}.
\end{equation}
Then, if required, individual $l$ degree portions $\mathbf{x}_{1:L}$ may be obtained from $\mathbf{x}_{1:L}$. Note that to avoid aliasing in signal reconstruction, a high enough $l$ degree truncation is required so that the high-frequency components of the signal do not get projected inaccurately onto basis vectors corresponding to low frequencies. It was shown in \cite{taulu} that a truncation at approximately $L=8$ is sufficient to represent signal vectors; in this paper, we truncate at $L=12$ in anticipation of close sensor array distances capturing signals of higher spatial complexities. Also note that $l\neq 0$ necessarily due to the absence of magnetic monopoles.


\section{Results}

\subsection{Simulation setup}

For our reference set-up, we used a simple 1-shell spherical head model of radius 9~cm and conductivity 0.33 S/m with origin located at the center of the sphere, and did a CC BEM forward calculation as described in Sections \ref{sec:BEM2} and \ref{sec:BEM3} to obtain $\bm{\phi}$. The BEM method was implemented using the Matlab library as provided in \cite{stenroosog}. The spherical surface was triangulated into 1280 triangles (642 vertices), and 4 dipole sources with varying depths located at $(2,0,0)$~cm, $(4,0,0)$~cm, $(6,0,0)$~cm, and $(8,0,0)$~cm were specified. All the dipoles had a moment of $(0, 10, 0)$~nAm. To simulate inaccurate mesh modeling to obtain $\bm{\phi}'$, the vertices of the spherical mesh were randomly perturbed radially up to 2\%, 4\%, 6\%, 8\% and 10\% relative to the 9~cm radius.

\begin{figure}[ht] 
    \centering
    \includegraphics[width=0.5\textwidth]{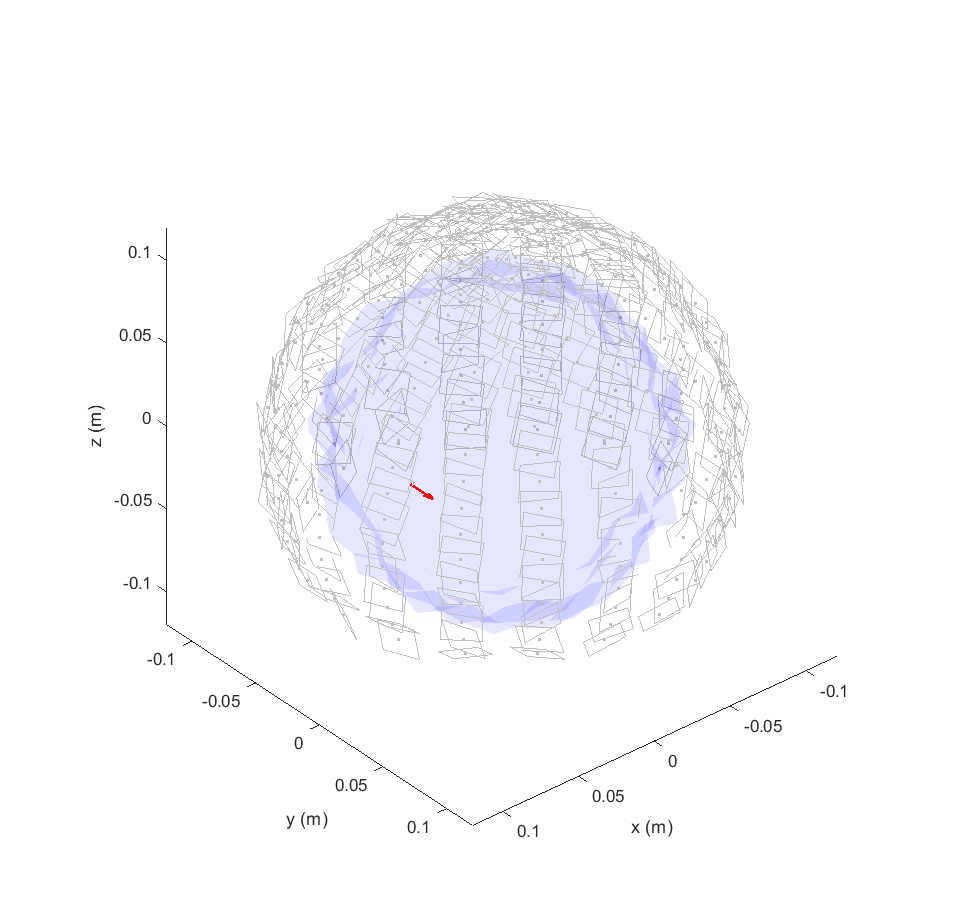}
    \caption{A $(6,0,0)$~cm dipolar source with moment $(0, 10, 0)$~nAm (red arrow) is shown located within a triangulated sphere of radius 9~cm that has its vertices randomly perturbed by up to 10\% (blue mesh). The spherical sensor array of radius 10~cm with 324 square pick-up loops of side length 2.1~cm and random orientations is also shown in gray.}
    \label{fig: setup}
\end{figure}

The sensor array that we used consisted of 324 square magnetometer pick-up loops with side length 2.1~cm all with non-radial orientations (to avoid linear dependence of the SSS basis \cite{taulu}), uniformly arranged on a spherical shell up to $\pi/6$ below the $z=0$ plane. It has been shown in \cite{yeo} that a 9-point cubature approximation yields accurate calculations for the sensor distances of 10~cm to 15~cm that we considered, thus we used a 9-point cubature approximation for the calculation of signal vectors in this paper. Figure~\ref{fig: setup} illustrates the sensor, head, and source setup as described above.

The averaged results over 100 forward calculations with this random vertex perturbation setup is presented in the following sections. 

\subsection{Signal vector error for sensor arrays at varying distances}

\begin{figure}[ht] 
    \centering
    \includegraphics[width=\textwidth]{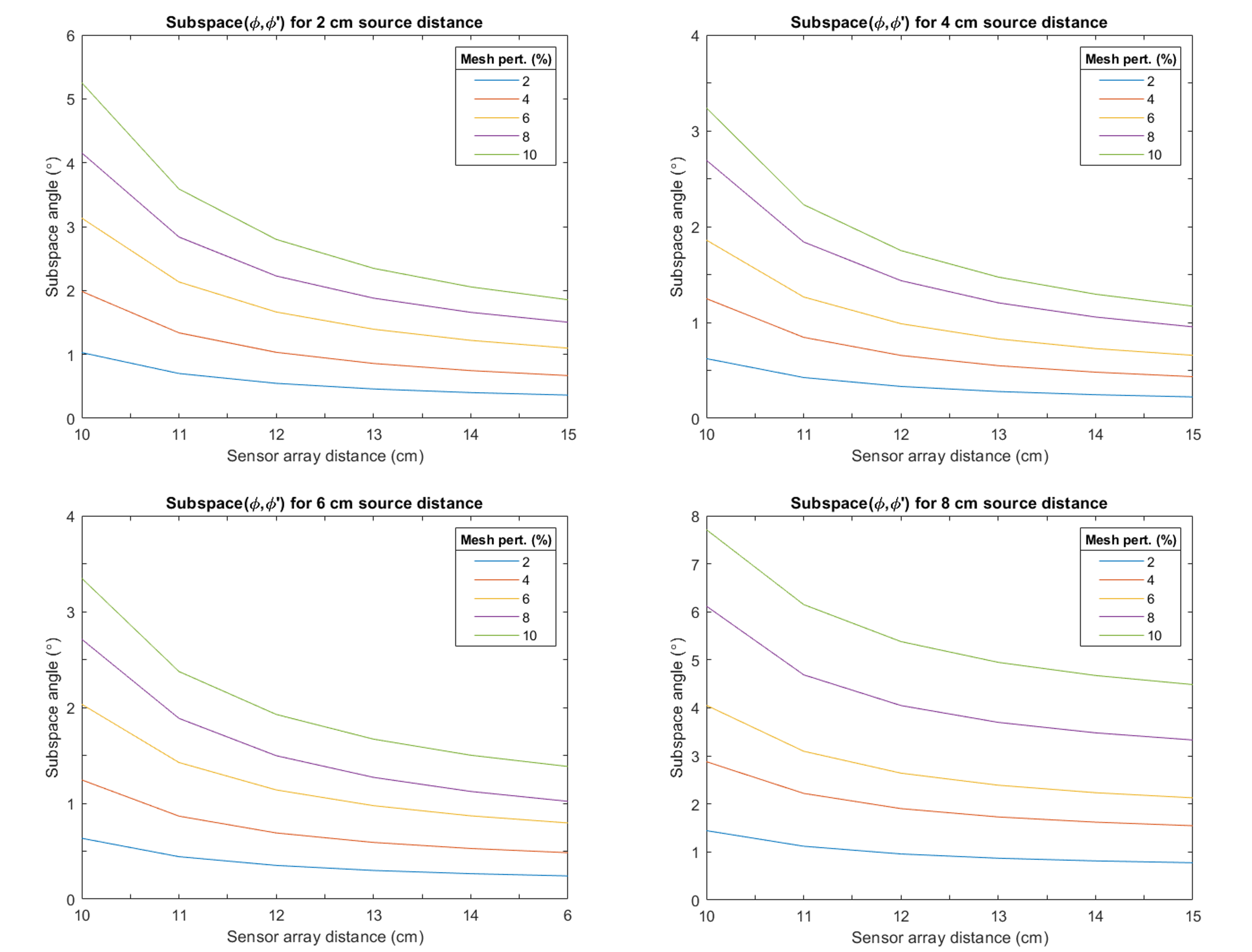}
    \caption{In the noiseless case, the decreasing subspace angle between the reference signal and perturbed signal as distance increases shows that the signal error caused by head model errors are less impactful for more distant sensor arrays. Intuitively, as the perturbation of the head model increases, the subspace angle increases as well.}
    \label{fig: err_all}
\end{figure}

\begin{figure}[ht] 
    \centering
    \includegraphics[width=\textwidth]{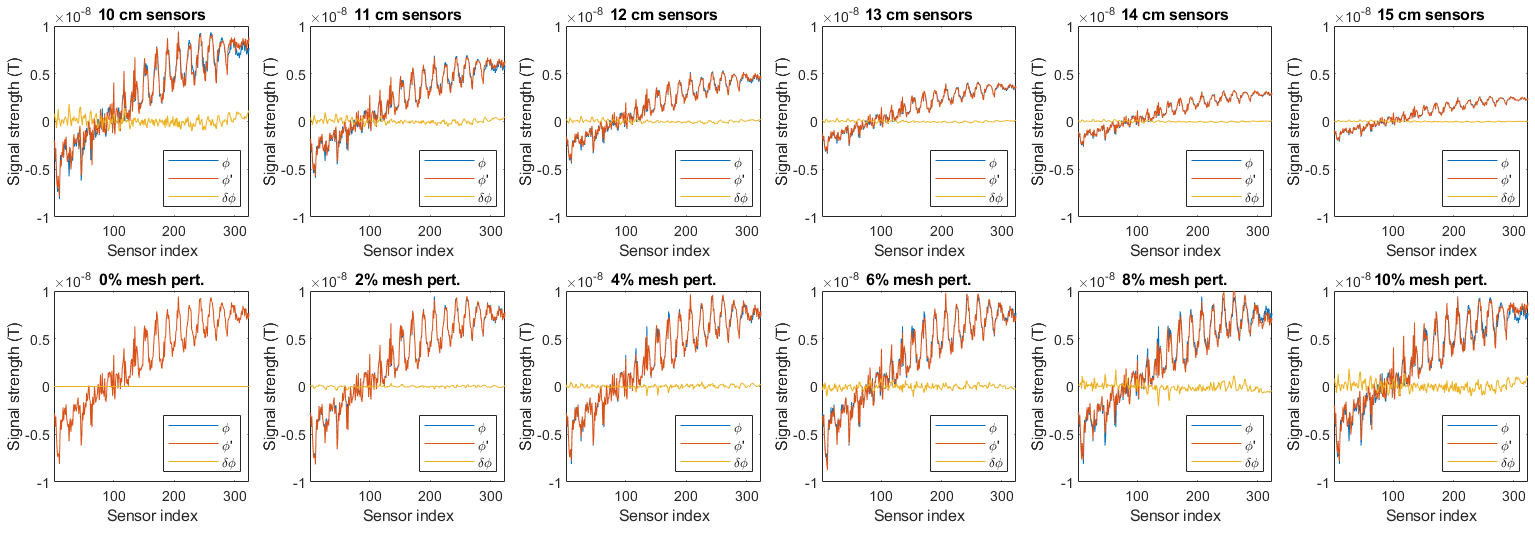}
    \caption{(First row) We plot the signal vectors $\bm{\phi}$ and $\bm{\phi}'$ and signal error $\bm{\phi}'-\bm{\phi}$ for the 2~cm source with a 10\% mesh perturbation across various sensor array distances. We see that indeed, as sensor array distance increases, signal error decreases for the same amount of mesh perturbation. (Second row) The 2~cm source with a sensor array distance of 10~cm was considered across the various mesh perturbations. We see that increased mesh perturbations give increased signal error.}
    \label{fig: abssig_err}
\end{figure}

First, we investigate how the error due to BEM mesh inaccuracy varies according to sensor array distance using equation \eqref{angle}. The sensor array radii were set to be from 10 cm to 15 cm, in increments of 1~cm (i.e. 1~cm to 6~cm from the surface of a 9~cm head model) and the signal was assumed the be noiseless. Figure~\ref{fig: err_all} shows that for all the source distances considered, as sensor array distance increases, the subspace angle between the reference signal $\bm{\phi}$ and perturbed signal $\bm{\phi}'$ decreases, indicating decreasing relative effects of mesh boundary inaccuracies as sensor array distance from the head increases. Moreover, smaller perturbations to the head model resulted in smaller subspace angles for a given sensor array distance when compared to higher perturbations as expected. These results may also be visually seen via plots of $\bm{\phi}$ and $\bm{\phi}'$ explicitly; we show this in Figure~\ref{fig: abssig_err} with the 2~cm source case across the various sensor distances and mesh perturbations for one of the 100 random mesh perturbation realizations.

\begin{figure}[ht] 
    \centering
    \includegraphics[width=0.6\textwidth]{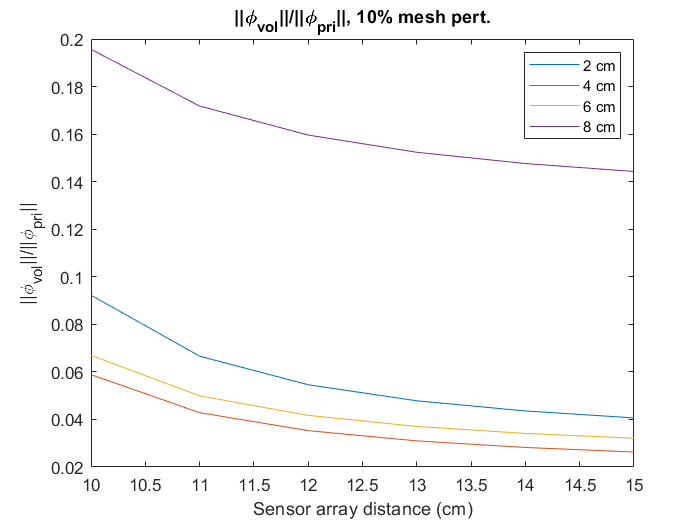}
    \caption{The volume-to-primary current contributions towards the total signal for all mesh perturbations is the highest for the 8~cm and 2~cm source distance cases; the 10\% mesh perturbation case is shown here. This implies highest signal errors at superficial and deep source locations for constant mesh perturbations, which agrees with the result in Figure~\ref{fig: err_all}.}
    \label{fig: volpri_meshpert}
\end{figure}

An interesting observation is that from the 2~cm source case to the 4~cm and 6~cm source cases, the subspace angle decreases before increasing again in the 8~cm source case; i.e. there is a ``turnaround'' point in subspace angle as a function of source distance. This may be explained by the fact that for deeper and superficial sources, the ratio between the volume current contribution to the primary current contribution is higher than for sources that lie between them. For all mesh perturbation cases, 4~cm and 6~cm sources have the lowest volume-to-primary current contributions to the signal as compared to the 2~cm and 8~cm sources. Figure~\ref{fig: volpri_meshpert} illustrates this for the 10\% mesh perturbation case. This may be explained qualitatively by the fact that the 2~cm deep primary source has a low signal strength due to its larger source-to-sensor distance, and hence has a low volume-to-primary current contribution to the signal. The superficial 8~cm source has a small source-to-sensor distance, and hence measures a larger volume current contribution, resulting in the higher volume-to-primary current contribution to the signal. As mentioned previously, all signal errors result from errors in the volume contribution portion only, i.e. $\delta \phi = \delta \bm{\phi}_{vol}$. Thus, the higher volume contribution relative to the primary contribution implies higher signal errors given a same amount of mesh perturbation (equivalently, volume current perturbation) for the 2~cm and 8~cm source cases, hence explaining the ``turnaround'' point of the subspace angles at some source distance in between them.

\begin{figure}[ht] 
    \centering
    \includegraphics[width=\textwidth]{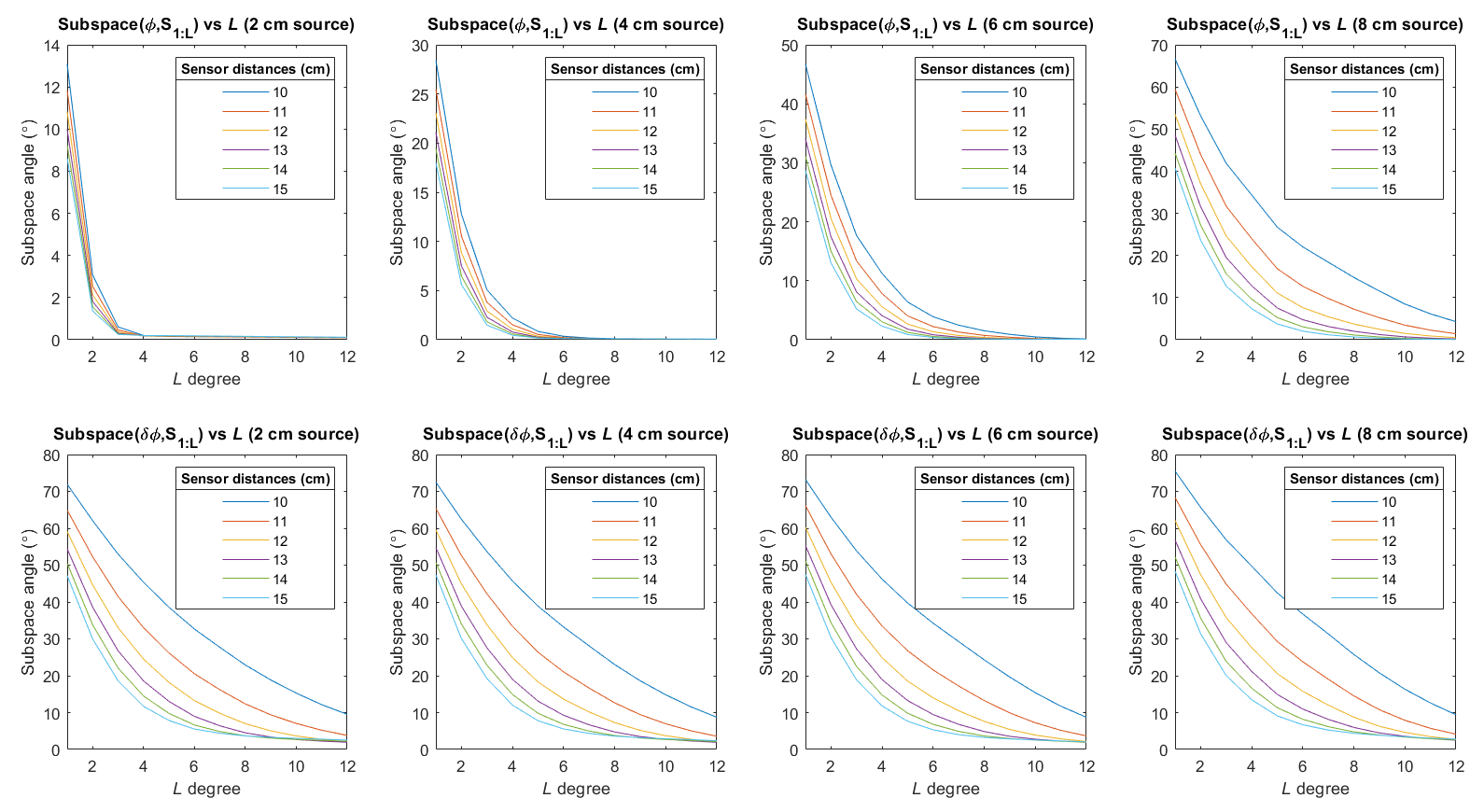}
    \caption{(Top row) The total signal $\bm{\phi}$ can be explained well at $L=12$ degree truncation of the VSH basis matrix $\mathbf{S}_{1:L}$ for all source distances and a 10\% mesh perturbation. For closer sensor distances, the subspace angle between $\bm{\phi}$ and $\mathbf{S}_{1:L}$ is higher given an $L$ degree truncation. (Bottom row) Similarly, for a given $L$ degree truncation and a given perturbed head model (10\% perturbation in this case), the subspace angle decreases for increasing sensor distances. However, the angles are higher than in the total signal case, indicating higher spatial complexity of the signal and signal errors recorded by sensor arrays closer to the head.}
    \label{fig: err_ldeg}
\end{figure}

Next, we determine if the additive error of the signal vector, $\delta \bm{\phi}$, may be explained by increasing orders of $l = L$ degree truncation, and if $\delta \bm{\phi}$ varies according to sensor distances. First, we show that the unperturbed signal may be explained to a large extent with $L=12$ degree truncation for all the sensor array distances that we considered. This is shown in the first row of plots in Figure~\ref{fig: err_ldeg}; the subspace angle with $\textbf{S}_{1:L}$ decays to become nearly zero at $L=12$ for all source distances. As expected, closer sensor array distances have higher subspace angles than further sensor arrays, which agrees with Figure~\ref{fig: err_all}.

From the second row of plots in Figure~\ref{fig: err_ldeg} however, we see that despite the total signal $\bm{\phi}$ being well-explained at $L=12$, the signal errors $\delta \bm{\phi}$ which consist mostly of higher frequency components require a higher $L$ truncation to explain the signal for all source distances. The subspace angle still decreases for increased $l$ degree truncation as expected (since $\delta \phi = \delta \phi_{vol}$ are still volume current contributions), and the subspace angles are smaller for increased sensor array distances. However, the angles are much higher than those of the total signal $\bm{\phi}$, indicating that higher spatial frequency components are affected more by inaccurate mesh modeling. This is because a higher $l$ degree truncation required to fully explain the additive signal error $\delta \bm{\phi}$ of closer sensor arrays indicates that their $\delta \bm{\phi}$ has a more spatially complex pattern with higher spatial frequency components. This higher spatial complexity is due to the fact that higher spatial frequency components decay quickly with increasing distance, hence by placing sensors closer to the source, these components can be captured to give higher resolution of the signal error $\delta \bm{\phi}$. Note that Figure~\ref{fig: err_ldeg} is for the case where the vertices of the mesh was perturbed radially up to 10\%. The plots for the different mesh perturbations are similar, hence we present just the 10\% case here.

\subsection{Source localization and orientation errors}

From our results above which considers the noiseless signal case, the signal vector suffers from higher inaccuracies for closer sensor arrays due to the larger errors in their higher spatial frequency components. Deep and superficial sources also had higher signal inaccuracies due to a higher volume current contribution relative to primary currents. Here, we investigate if these observations will be seen in the form of source localization errors as well. The source localization procedure was done via a standard ECD fit using the ``fit\_dipole'' function in MNE-Python~1.0 \cite{GramfortEtAl2013a,GramfortEtAl2014}. The forward model here was computed using LC BEM however, as it is currently the only solving method implemented in MNE-Python.

\begin{figure}[ht] 
    \centering
    \includegraphics[width=\textwidth]{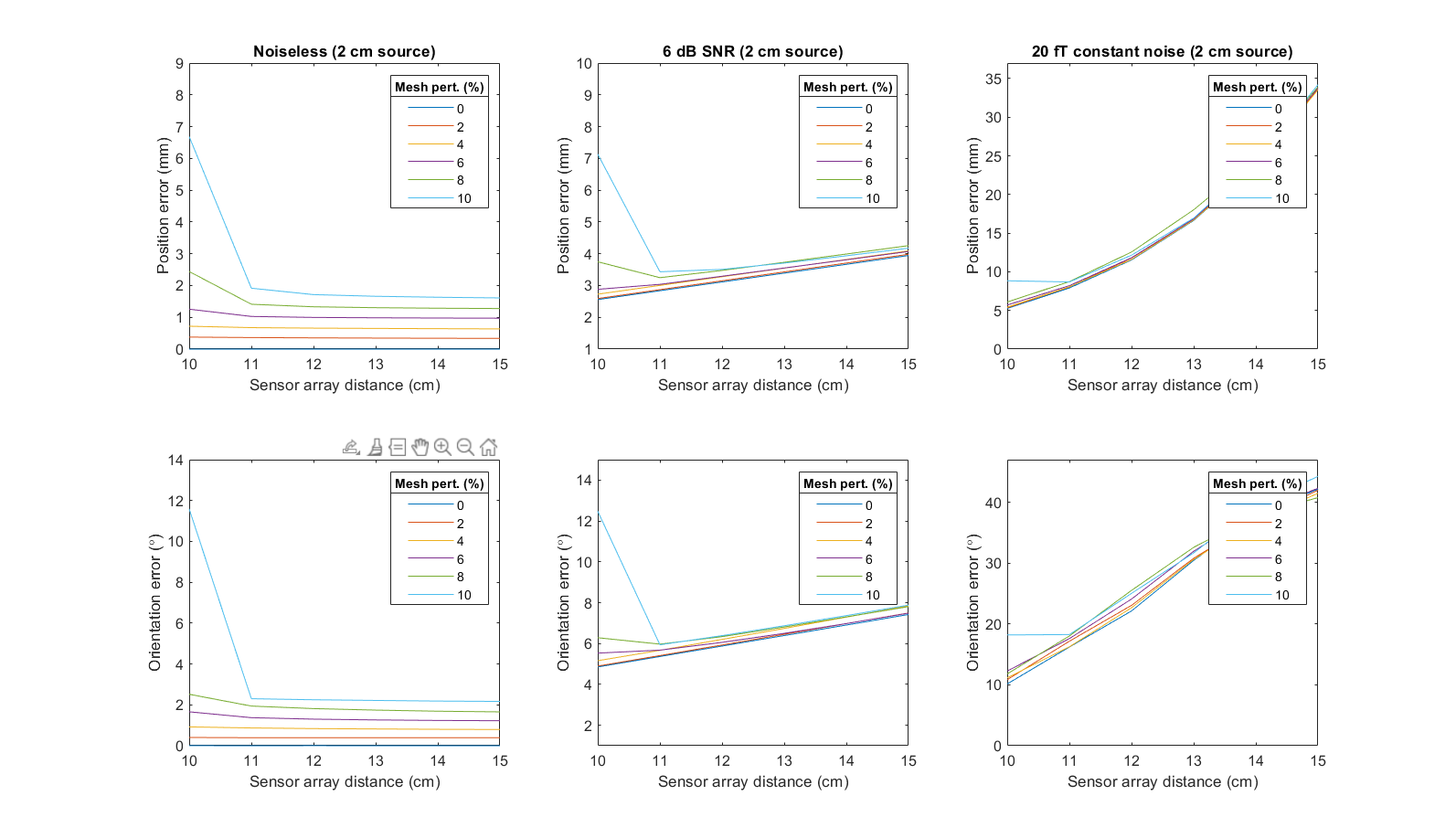}
    \caption{2~cm source case. (Left column) The position and orientation errors of the ECD fit for the noiseless signal case is shown. For closer sensor array distances, the error increases. This agrees with the result that closer sensor array measurements give more inaccurate signals in the noiseless case. (Middle column) In the noisy case where we have a SNR level of 6~dB, the localization and orientation errors start to increase slightly as sensor array distances increase, indicating that the effects of better localization due to higher SNR is starting to balance out the effects of poorer localization due to more inaccurate signals captured by closer sensor array distances. (Right column) In the noisiest case with a constant noise level of 20 fT, the dipoles are better localized at closer sensor distances since the effect of higher SNR now outweighs the effects of signal error due to head model inaccuracies.}
    \label{fig: err_loc2}
\end{figure}

\begin{figure}[ht] 
    \centering
    \includegraphics[width=\textwidth]{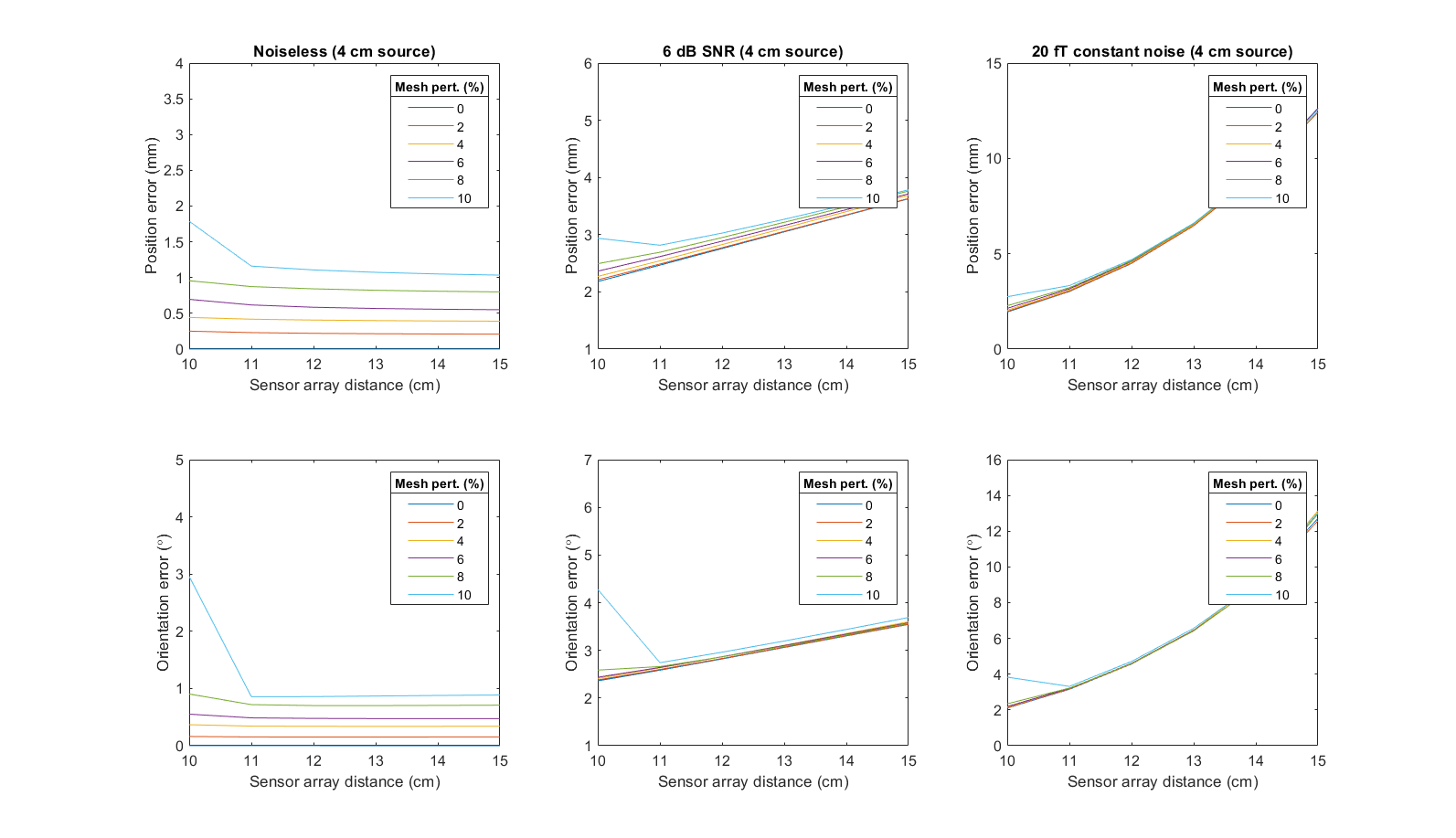}
    \caption{4~cm source case. The description of this figure is similar to Figure~\ref{fig: err_loc2}.}
    \label{fig: err_loc4}
\end{figure}

\begin{figure}[ht] 
    \centering
    \includegraphics[width=\textwidth]{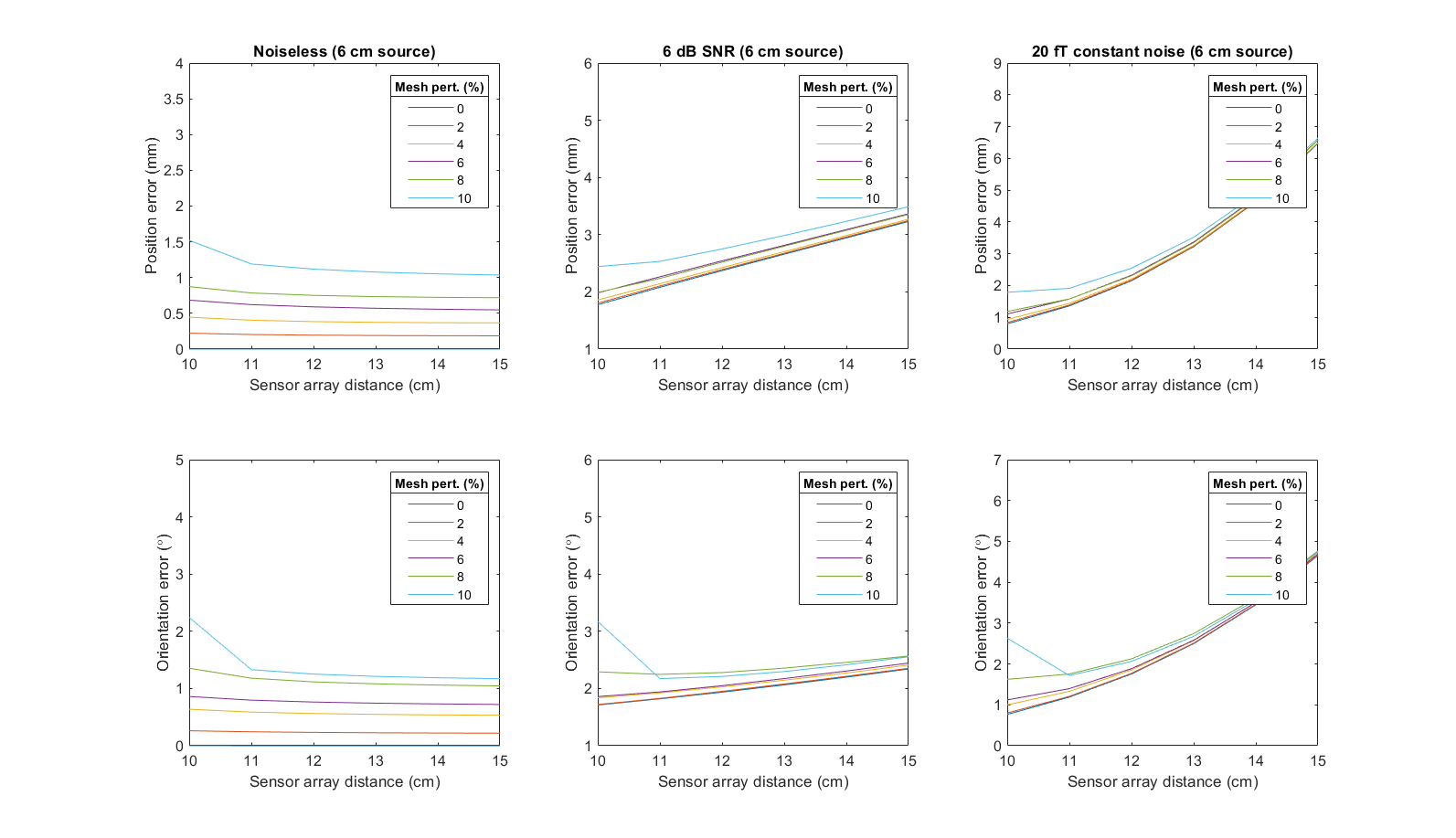}
    \caption{6~cm source case. The description of this figure is similar to Figure~\ref{fig: err_loc2}.}
    \label{fig: err_loc6}
\end{figure}

\begin{figure}[ht] 
    \centering
    \includegraphics[width=\textwidth]{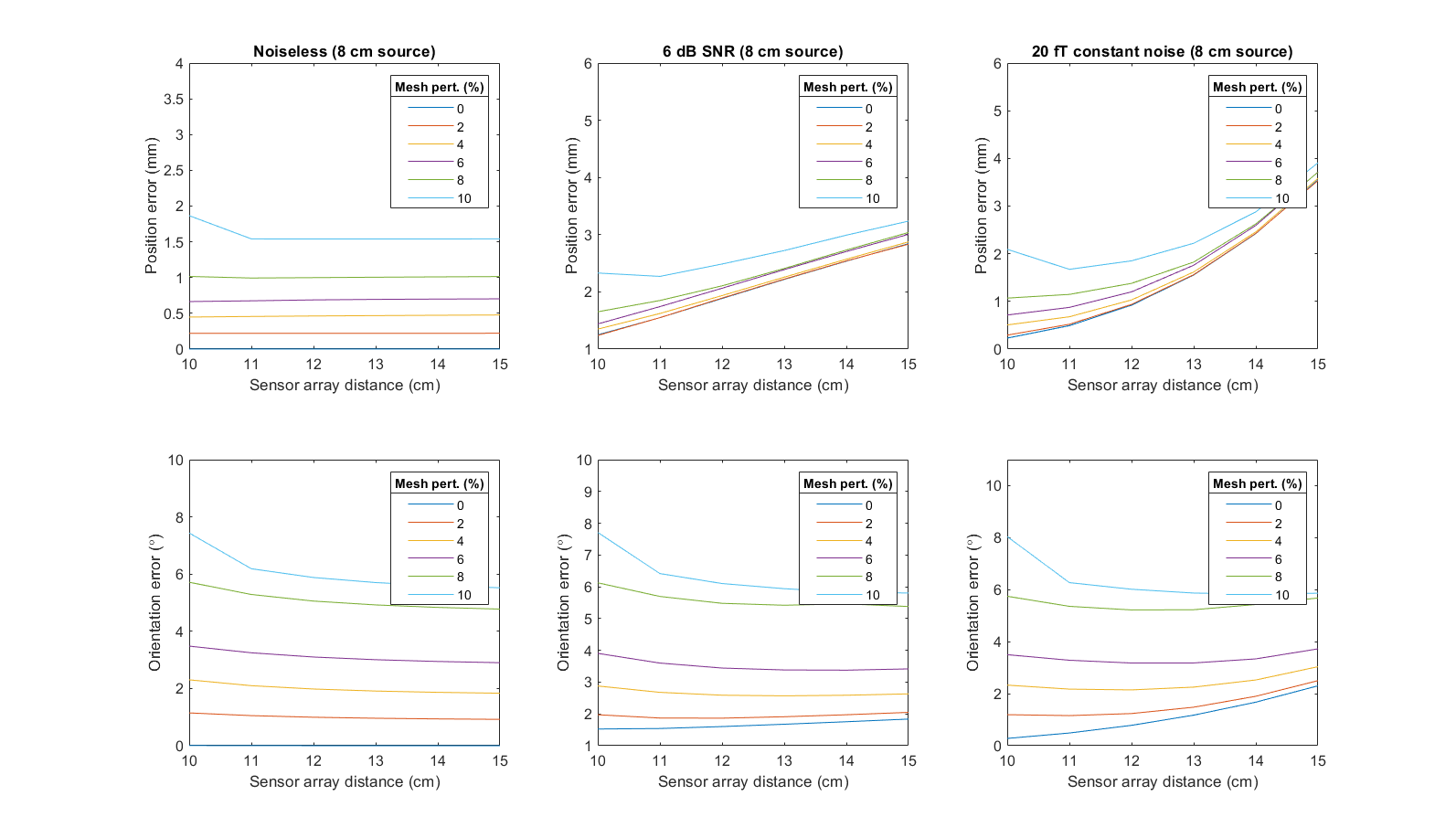}
    \caption{8~cm source case. The description of this figure is similar to Figure~\ref{fig: err_loc2}.}
    \label{fig: err_loc8}
\end{figure}

The results for the noiseless case are shown in the first columns of Figures~\ref{fig: err_loc2}, \ref{fig: err_loc4}, \ref{fig: err_loc6} and \ref{fig: err_loc8}, which correspond to the 2~cm, 4~cm, 6~cm and 8~cm source case respectively. Indeed, in the noiseless case, source localization and orientation errors are largest for closer sensor arrays, as well as deep and superficial sources for a fixed mesh perturbation. These observations are in agreement with the result for signal errors as presented above.

However, the primary motivation to move sensors closer to the head is because closer sensors have the potential to give more accurate source localization results due to higher SNR as mentioned in the introduction. The definition of SNR (with units of decibels) used in MNE-Python~1.0 follows that of Goldenholz \cite{goldenholz},
\begin{align}
    \text{SNR} = 10 \log \left\lfloor \frac{q^2}{N} \sum_{j=1}^N \frac{\phi_j^2}{s_j^2} \right\rfloor
\end{align}
where $q$ is the source strength, $N$ is the total number of sensors on the sensor array, $\phi_j$ is the signal on sensor $j$ and $s_j^2$ is the noise variance on sensor $j$. We thus introduced various noise levels to determine the noise level at which the effect of having more accurate source localization due to higher SNR of closer sensors outweighs the effect of less accurate source localization due to the higher errors of the high frequency components of the noiseless signal. We found that this occurred at a SNR of around 6~dB (the sensor noise was varied to force a 6~dB SNR for a varying sensor array distances); the localization and orientation errors began to increase as sensor array distances increased at around an SNR of 6~dB. This is shown in the second column of Figures~\ref{fig: err_loc2}, \ref{fig: err_loc4}, \ref{fig: err_loc6} and \ref{fig: err_loc8}. The rightmost column of Figures~\ref{fig: err_loc2}, \ref{fig: err_loc4}, \ref{fig: err_loc6} and \ref{fig: err_loc8} show that for an SNR greater than 6~dB, in this case a constant 20~fT noise level, then there is an improvement in source localization for closer sensor arrays. This means that a higher SNR allowing for better source localization resolutions outweighs the effects of BEM errors for noisy signals with SNR above approximately 6~dB.

\section{Conclusion}

In this paper, we have investigated the signal and source localization errors due to BEM head geometry inaccuracies with respect to varying MEG sensor array distances. This examination was motivated by next-generation OPM sensors that may be placed directly on the subject's head, resulting in signal measurements with higher spatial resolutions. This includes measuring any inaccurate volume current contributions with higher resolution; as stated before, in a piecewise homogeneous conductor model of the head, volume currents may be equivalently expressed as surface currents at the head model boundaries. As such, head model inaccuracies lead to inaccurate volume current contributions.

We found that for signals with SNR greater than 6~dB, placing sensor arrays closer to the head causes higher relative signal errors for a perturbed spherical head model, and thus this leads to higher source localization errors. This is because the signal measured by these sensor arrays now contain finer spatial details due to the ability to now capture the fast-decaying high spatial frequency components of the magnetic field, and these high frequency components suffer from higher errors due to head model inaccuracies. Moreover, for deep and superficial sources where there are higher relative volume current contributions to the signal, the signal and source localization errors are higher as well for a fixed mesh perturbation amount. For signals with SNR below 6~dB, the advantage of having higher SNR for closer sensor array distances outweighs the effects due to the increased errors arising from the high spatial frequency components, thus source localization errors decrease; this is consistent with the current understanding of the advantages of using OPM sensor arrays.

Our results tell us that in general, for noisy signals, OPM sensors should indeed result in more accurate source localization results despite head modelling errors. However, as signal noise levels decrease either via an improvement in sensor technology or signal processing methods (e.g. averaging over many trials), BEM head geometry errors must be minimized when sensor distances are shifted closer to the head in order to avoid increased signal and source localization inaccuracies.

\section{Acknowledgements}
We would like to thank Matti Stenroos for helpful discussions regarding BEM, as well as clarifications about his BEM codes. This project was funded by the NIH Brain Initiative grant U01EB028656 awarded to Sandia National Laboratories.

Sandia National Laboratories is a multimission laboratory managed and operated by National Technology and Engineering Solutions of Sandia, LLC—a wholly owned subsidiary of Honeywell International Inc.—for the U.S. Department of Energy’s National Nuclear Security Administration, under contract DENA0003525. This paper describes objective technical results and analysis. Any subjective views or opinions that might be expressed in the paper do not necessarily represent the views of the U.S. Department of Energy, the United States Government, or the National Institutes of Health. The content is solely the responsibility of the authors.

\appendix

\section{Analytical first-order errors for CC BEM} \label{appendix_analytical}

From \eqref{Bfield_CC}, we see that there are three terms that depend on the head model accuracy that affect the calculation of the magnetic field: the conductivities $\sigma_{l}^{\pm}$, the scalar electric potential $V(\mathbf{c}_l^m)$ (equivalently, the matrix $\mathbf{G}$), and the vector solid angle $\mathbf{\Omega}_l^m$. The latter two depend on the mesh accuracy, whereas the former depends on the conductivity values.

\subsection{Perturbations to mesh vertices}

First, we consider perturbations to $V$. By looking at the form of $\mathbf{G}$ as in \eqref{G}, we see that there are two possible sources of error in the head model: the solid angle (i.e. vertex/triangle centroid perturbations), and the conductivity values of each region. Here, we offer analytical forms to compute the first-order errors due to each of these sources of error.  Higher order corrections may be obtained with the help of e.g. Matlab or Mathematica.

The solid angle can be regarded as a scalar field in 12-dimensional space (3 coordinates per $\mathbf{r}$, $\mathbf{r}_1$, $\mathbf{r}_2$, $\mathbf{r}_3$). A perturbation to one of these 12 coordinates corresponds to perturbations in the $x$, $y$ or $z$ direction of one of the three vertices of the corresponding triangle $\mathbf{r}_1$, $\mathbf{r}_2$, $\mathbf{r}_3$, or the $x$, $y$ or $z$ direction of the point of reference $\mathbf{r}$. Let $\mathbf{r} = (x,y,z)$ and $\mathbf{r}_i = (x_i,y_i,z_i)$, where $i = 1,2,3$. For small perturbations, only the first-order expansion term is significant. If we let $x$ be perturbed to become $x + \delta x$, the error in solid angle calculations \eqref{solidangle}, $\delta \Omega$, can be approximated as
\begin{equation} \label{dOmega_fo}
    \delta \Omega = \Omega \left(x+\delta x, y, z, x_1, \dots, z_3 \right) - \Omega \left(x, \dots, z_3 \right) \approx \left. \frac{\partial \Omega}{\partial x}\right|_{\mathbf{r},\mathbf{r}_{1},\mathbf{r}_{2},\mathbf{r}_{3}} \delta x
\end{equation}
Let the argument within the arctan be $P = P(x,\dots,z_3)$. The right hand side of \eqref{dOmega_fo} may be evaluated by the chain rule:
\begin{equation}
    \frac{\partial \Omega}{\partial x} = \frac{\partial \Omega}{\partial P} \frac{\partial P}{\partial x} = \frac{2}{1+P^2} \frac{\partial P}{\partial x}.
\end{equation}
Note that due to the symmetric form of the solid angle where the numerator obeys scalar triple product identity and the denominator has all terms that obey cyclic index permutations, we only need to evaluate 2 partial derivatives of $P$ to get all 12 of them. Namely, we only need to evaluate one partial derivative with respect to any of the 3 coordinates of $\mathbf{r}$, and another with respect to any of the 9 coordinates of $\mathbf{r}_i$. Cyclic permutation of the coordinate indices $(x,y,z) \leftrightarrow (z,x,y) \leftrightarrow (y,z,x)$ then gives us the other partial derivatives with respect to the other 2 coordinates of the corresponding $\mathbf{r}$ or $\mathbf{r}_i$. Then, for the case of partial derivatives corresponding to coordinates of $\mathbf{r}_i$, permutation of vertex indices $(1,2,3) \leftrightarrow (2,3,1) \leftrightarrow (3,1,2)$ gives us the other 2 triangle vertices' 6 partial derivatives. Note that if we extend to higher-order partial derivatives, symmetry considerations may still be used to reduce the total number of partial derivatives to evaluate. However, mixed partials mean that more than 2 partial derivatives need to be evaluated necessarily. 


One may also interpret the above 12 partial derivatives to evaluate as a 1-1 correspondence between the index permutations and 12 coordinates. The 3 $\mathbf{r}$ coordinates correspond to the 3 cyclic permutations on the coordinate indices $(x,y,z)$, whereas the 9 $\mathbf{r}_i$ coordinates correspond to the $3\times 3 = 9$ possible pairings of the cyclic permutations of two sets of indices, namely the coordinate indices $(x,y,z)$ and vertex indices $(1,2,3)$. If we denote the 3 coordinates of $\mathbf{r}$ as $(x_1,x_2,x_3)$ and the 9 coordinates of $\mathbf{r}_i$ as $x_j$, $j = 4, \dots, 12$, then the total perturbation of the solid angle is
\begin{equation}\label{dOmega}
    \delta \Omega_l^m \left(\mathbf{c}_k^i \right) \approx \frac{2}{1+P^2} \left(\sum_{j=1}^3  \left. \frac{\partial P}{\partial x_1} \right|_{\{\sigma^c_j\}} \delta x_j + \sum_{j=4}^{12} \left. \frac{\partial P}{\partial x_4} \right|_{\{\sigma^{c+v}_j\}} \delta x_j\right)
\end{equation}
where $\sigma^c_j = (x_j,x_k,x_l)$ are the 3 possible coordinate index cyclic permutations, and $\sigma^{c+v}_j$ are the 9 possible pairs of coordinate index and vertex index cyclic permutations. Any small perturbation of a vertex results in the adjacent triangles' perturbed centroids (i.e. perturbed observation points $\mathbf{r}$) and the adjacent triangles' change in solid angles; these two cases correspond to the first and second sums in \eqref{dOmega} respectively.


The effects of the perturbations above may be represented by a sparse additive perturbative matrix $\delta \mathbf{G}$ as defined in \eqref{dG}, whose elements are calculated by \eqref{dOmega} and \eqref{dsigma}. Note that the first sum of \eqref{dOmega} corresponding to the case of perturbed centroids contribute to nonzero row entries, whereas the second sum corresponding to perturbed vertices contribute to nonzero column entries, due to our arrangement of the block elements in $\mathbf{G}$.

We now want to see how this affects $\mathbf{V}$. Let $\mathbf{A} \equiv (\mathbb{I}-\mathbf{G} + \mathbf{e}\mathbf{c}^T)$ and $\tilde{\mathbf{A}} \equiv \mathbf{A} + \delta \mathbf{G}$. If $\tilde{\mathbf{A}}$ is non-singular, then \cite{stewart} 
\begin{align}
    \tilde{\mathbf{A}} \tilde{\mathbf{A}}^{-1} &= \left(\mathbf{A} + \delta \mathbf{G}\right)\tilde{\mathbf{A}}^{-1} = I \\
    \implies \mathbf{A}^{-1} &= \left( I + \mathbf{A}^{-1} \delta \mathbf{G}\right)\tilde{\mathbf{A}}^{-1} \\
    \implies \tilde{\mathbf{A}}^{-1} - \mathbf{A}^{-1} &= - \mathbf{A}^{-1} \delta \mathbf{G} \left(\mathbf{A}+\delta \mathbf{G}\right)^{-1}
\end{align}
Therefore, errors in potential are given by 
\begin{equation}\label{V_perturb}
    \delta \mathbf{V} = -\mathbf{A}^{-1} \delta \mathbf{G} \left(\mathbf{A}+\delta \mathbf{G}\right)^{-1} \mathbf{V}_\infty.
\end{equation}

Next, we consider the first-order perturbation to the vector solid angle \eqref{vecomega}. It depends only on the coordinates of the triangles' vertices and may be obtained in a straightforward manner,
\begin{equation}\label{vecomega_perturb}
    \delta \mathbf{\Omega}_l^m \approx \sum_{i=1}^{3} \left[ \sum_{j=1}^{9} \frac{\partial\left(\gamma_{i-1} - \gamma_i \right)}{\partial x_j} \delta x_j \right]\mathbf{r}_i.
\end{equation}

\subsection{Perturbations to conductivity}

We now consider perturbations to $\sigma^{\pm}$, which are conductivity values within the layers of the head model. This is an easier case to deal with, since we may simply add a perturbative constant $\delta \sigma$ to each $\sigma$. For the conductivity term within $\mathbf{G}$, let us denote this error term as
\begin{align} \label{dsigma}
    \delta \sigma_{k,l}  &= \frac{\sigma_l^- + \delta \sigma_l^- - \sigma_l^+ - \delta \sigma_l^+}{\sigma_k^- + \delta \sigma_k^- + \sigma_k^+ + \delta \sigma_k^+} - \frac{\sigma_l^- - \sigma_l^+}{\sigma_k^- + \sigma_k^+ }  \nonumber \\
    &= - \frac{\left(\sigma_l^- - \sigma_l^+ \right)\left(\delta \sigma_k^- + \delta \sigma_k^+ \right) + \left(\sigma_k^- + \sigma_k^+ \right)\left(\delta \sigma_l^+ - \delta \sigma_l^- \right)}{\left(\sigma_k^- + \sigma_k^+ \right)^2 + \left(\sigma_k^- + \sigma_k^+ \right) \left(\delta \sigma_k^- + \delta \sigma_k^+\right) } 
\end{align}
Together with \eqref{dOmega}, the total perturbative matrix $\delta \mathbf{G}$ for small vertex perturbations and arbitrary conductivity inaccuracies is
\begin{equation} \label{dG}
    \delta G_{k,l}^{i,m} =  - \frac{1}{2\pi} \delta \sigma_{k,l} \delta \Omega_l^m \left(\mathbf{c}_k^i \right).
\end{equation} 
The errors in the total magnetic field up to first order with respect to perturbations of the components of $\mathbf{r}_i$ as well as conductivities $\sigma$ may be given by
\begin{align} 
    \delta \mathbf{B}\left(\mathbf{r} \right) \approx \frac{\mu_0}{4 \pi} \left\{ \sum_{l=1}^{N_S} \left(\sigma_l^- - \sigma_l^+\right) \sum_{m=1}^{N_l} \left[ \delta V\left(\mathbf{c}_l^m \right)\mathbf{\Omega}_l^m + V\left(\mathbf{c}_l^m \right) \delta \mathbf{\Omega}_l^m \right] + \sum_{l=1}^{N_S} \left(\delta \sigma_l^- - \delta \sigma_l^+\right) \sum_{m=1}^{N_l} V\left(\mathbf{c}_l^m \right)\mathbf{\Omega}_l^m \right\}
\end{align}
with its terms given by \eqref{V_perturb}, \eqref{dG}, \eqref{dsigma}, \eqref{dOmega}, and \eqref{vecomega_perturb}.

\end{document}